\documentclass[12pt]{iopart}
\usepackage{graphicx}

\begin{document}
\title[Loss mechanisms of a 3D superconducting microwave cavity for dark matter searches]{Revealing the loss mechanisms of a 3D superconducting microwave cavity for use in a dark matter search}

\author{J. C. Esmenda$^1$, E. A. Laird$^1$, I. Bailey$^1$, N.\,Du$^2$, S.\,Durham$^2$, G. Carosi$^2$, T. Gamble$^3$, P. Smith$^3$, E. Daw$^3$ and Yu. A. Pashkin$^1$}

\address{$^1$ Department of Physics, Lancaster University, Lancaster, LA1 4YB, UK}
\address{$^2$ Lawrence Livermore National Laboratory, Livermore, California 94550, USA}
\address{$^3$ School of Mathematical and Physical Sciences, University of Sheffield, Sheffield, S3 7RH, UK}
\ead{j.esmenda@lancaster.ac.uk}
\ead{y.pashkin@lancaster.ac.uk}

\begin{abstract}
Superconducting microwave cavities have found applications in many areas including quantum computing, particle accelerators, and dark matter searches. Their extremely high quality factors translate to very narrow bandwidth, which makes them key components of sensitive detectors. In this study, we aim to understand the loss mechanisms of an aluminium cavity and how they change as the cavity material transitions from the superconducting to normal state. We found that at temperatures not much lower than the transition temperature $T_c$, losses are dominated by quasiparticle excitations and are well described by the BCS theory. The exponential decrease of the quasiparticle density below $T_c$ results in a 1000-fold increase of the quality factor, as well as a shift of the resonance frequency due to the change of the kinetic inductance of the superconductor. At very low temperatures, losses due to two-level systems begin to dominate giving a peak in the quality factor of about $2.76\times10^7$ at 130\,mK. Understanding the loss mechanisms is invaluable, as the working temperature of the cavity may vary during operation regardless of its application.
\end{abstract}

\vspace{2pc}

\submitto{\NJP}
\maketitle

\section{Introduction}
\label{sec:introduction}
The ability of a superconducting microwave cavity to retain a large amount of energy compared to how much power it loses per cycle is the fundamental basis of its value as a key component in a number of important research areas. These include particle physics, where the mature technology of superconducting radio-frequency (SRF) cavities is used for accelerators \cite{padamsee2014superconducting}, and quantum computing \cite{blais2004cavity,krasnok2024superconducting,reagor2016superconducting}, where long coherence times are vital for robust quantum information processing and error correction. This paper focuses on the application of such cavities to dark matter searches. The use of three-dimensional (3D) microwave cavities for light dark matter searches was first proposed \cite{sikivie1983experimental,sikivie1985detection} and realised \cite{depanfilis1987limits,hagmann1990results} in the 1980s, and continued by the ADMX collaboration at the start of the century \cite{asztalos2001large,asztalos2010squid}. In recent years a large number of other experiments of this kind have been proposed. The majority of these are axion haloscopes featuring a resonant microwave structure immersed in a strong, static magnetic field. This field enables the conversion of hypothetical axion dark matter to photons, and the most common resonant structure used to enhance the photon signal is a cylindrical cavity. Axion conversion is then searched for by measuring the power contained in the modes of the cavity, where the fundamental transverse magnetic mode (TM\textsubscript{010}) is typically the most sensitive. The requirement of an external magnetic field makes the use of superconducting cavities challenging, but the use of type II superconductors for this purpose has been demonstrated \cite{alesini2019galactic,golm2022thin}. A superconducting cavity can also be used to probe the cold dark matter paraphoton-photon coupling, as this does not require a magnetic field. Such experiments can either be sensitive to dark matter paraphotons using a haloscope geometry as in the axion case discussed above, or else both produce and detect hypothetical paraphotons using a ``light shining through a wall'' (LSW) geometry \cite{jaeckel2008cavity,parker2013hidden,povey2011microwave,caspers2009feasibility,povey2010microwave,parker2013cryogenic}. The advantage of superconducting cavities is due to the very low resistive loss the cavity wall achieves when the material turns superconductor, sustaining standing waves inside it before leaking energy. In other words, the ratio of the energy stored inside the cavity over the energy lost during one radian of oscillation, more commonly referred to as its quality factor, becomes extremely high. This then translates to very narrow bandwidths, which are an important feature of sensitive detectors. In practice, however, the expected high quality factor values are not always achieved. This suggests that there are other loss mechanisms that persist even when the cavity is superconducting \cite{martinis2005decoherence,gao2008physics,pappas2011two,zmuidzinas2012superconducting,krasnok2024superconducting}. For this reason, it is imperative to understand how these different loss mechanisms occur and behave if one is to maximise the efficacy of this technology.

To this end, we characterise an aluminium cavity with a cylindrical geometry, developed as part of the ADMX collaboration's experimental programme, in its superconducting temperature range. We then explore how losses due to surface imperfections in three-dimensional cavities are minimised with the aid of simulations. More importantly, we try to understand temperature dependent losses and how they contribute to the overall observed quality factor. While detailed studies of losses have been carried out on 3D superconducting cavities of different geometries \cite{reagor2013reaching,kudra2020high,romanenko2017understanding,romanenko2020three,bafia2021anomalous}, a careful study of the losses for this geometry is of particular interest to the dark matter search community. Moreover, the insights learned in this study will be an important addition to the knowledge base of superconducting cavities, in general.
\section{Cavity Characterisation}
\label{sec:characterisation}
The cavity, which is a cylinder with an interior length and radius of 121\,mm and 46.05\,mm, respectively, is formed from two mirrored halves of a milled Al 6061-T6 alloy with rounded edges as illustrated in Fig.~1a. To probe the cavity, two small through holes on one of the circular faces are made. It is important to note here that there are also two larger holes located on both circular faces of the cavity that are designed for a frequency tuning rod insertion but are not used for this experiment. Eight copper clamps tightened with brass screws embrace the cavity and are attached to the mixing chamber plate of a dilution refrigerator with a base temperature of 8\,mK as shown in Fig.~1b. Using a calculated thermal equilibration time of tens of milliseconds for a decimetre-sized volume of superconducting aluminium, a thermalisation time of at least 40\,minutes in between temperature points ensures that the temperature of the cavity is approximated well by the thermometer of the plate. Transmission measurements through the cavity are done by using two copper coaxial cables, where the central line at one end protrudes by a small amount to the probe through-holes while being secured from the outside using collet screws as shown in Fig.~1c. The insertion depth of the antenna is adjusted at room temperature such that the coupling to the external circuitry is minimised. It should also be noted here that upon receiving the cavity from the manufacturer, the cavity was cleaned modestly without any chemical treatment or additional machine polishing before any assembly and measurement were performed. Even with minimal preparations, the cylindrical cavity, which has a fundamental resonance frequency $f_0$ of 2.5123\,GHz, was measured to have a quality factor $Q$ of 26.095 $\times$ $10^{6}$ at the base temperature using a Lorentzian fit as shown in Fig.~1d.

\section{Effect of geometry on surface losses}
\label{sec:geometrical effect}
%
Finite element method simulators such as COMSOL \cite{comsol} allow for determining the frequencies and visualisation of the resonance modes of the cavity, and more importantly, can be used to understand the geometrical effect on losses. Table 1 summarises the measured resonance frequencies and quality factors of the different modes of the cavity. In the same table, the corresponding simulated frequencies are then calculated using the COMSOL model. The mode names in the first column of Table 1 indicate whether the mode is transverse electric (TE) or transverse magnetic (TM) while the succeeding numbers represent the nodal indices for the azimuthal, radial, and transverse degrees of freedom with the subscripts a and b designating modes with the same nodal indices but split due to asymmetry. Visualisation of the corresponding modal volumes of the electrical fields of these resonances shown in Fig.~2 help us understand why, for example, no peak is observed for the TE\textsubscript{101} mode. This is due to the transmission antenna being located in a low-field region of TE\textsubscript{101} and perpendicular to the electric field lines. This makes TM\textsubscript{010} the lowest observed mode in this experiment. It is also the mode of most interest to the majority of light dark matter searches. It is important to mention here as well that seam loss is minimised by using TM\textsubscript{010} mode in this case because the current flows in the same direction as the seam. To see the geometrical effect on losses, we begin by understanding surface imperfections.

Surface imperfections for a three dimensional cavity account for two types of losses, which ultimately determine its quality factor. The first type comes from the oxide layer at the surface, characterised by a dielectric loss tangent $\tan{\delta}$ or $1/\tan{\delta} = Q$\textsubscript{diel} in terms of $Q$. To estimate the actual $Q$ due to this loss, one needs to calculate the dielectric participation $P_{\mathrm{diel}}$. By summing up the electric field for the cavity's interior surface area $S$ and interior volume $V$ and taking the ratio, the dielectric participation can be calculated by the following equation:
\begin{equation}
P_{\mathrm{diel}} = \frac{\varepsilon_r t \int\int |\vec{E}|^2 dS}{\int\int\int |\vec{E}|^2 dV},
\end{equation}
where $t$ and $\varepsilon_\mathrm r$, are the thickness and relative permittivity, respectively. With this, surface oxide $Q$ will be $Q_{\mathrm{int,E}} = \rm Q_{diel}/P_{diel}$ \cite{pozar2011microwave}. Using COMSOL, the dielectric participation can be estimated for the different modes for an oxide thickness $t =$ 3\,nm as shown in Table 1. From these values, one can see that the dielectric participation values for the cavity modes of this geometry are very low. As an example, using the calucalted dielectric participation for TM\textsubscript{010}, $Q_{\mathrm{int,E}}$ can be as high as $10^{11}$ (for $\tan{\delta}$ = 0.0001) \cite{zmuidzinas2012superconducting}. This is further reinforced by the independence of the fundamental response peak on the input power as opposed to planar resonators, where the presence of oxide lead to power dependent frequencies and lifetimes \cite{martinis2005decoherence}. As shown in Fig.~1d, there was no observed change in $f_0$ and $Q$ between high power (-79\,dBm) and low power (-139\,dBm) at the cavity input port.

The second surface loss is due to the surface resistance $R_\mathrm s$, possibly from a finite quasiparticle density. Estimating the $Q$ due to this loss, one needs to determine the geometric factor $G$ ($\Omega$) \cite{gurevich2017theory,banford1961feasibility} which is given by:
\begin{equation}
G = \frac{\omega \mu_0 \int\int\int |\vec{H}|^2 dV}{\int\int |\vec{H}|^2 dS}
\end{equation}
for a mode angular frequency $\omega$, vacuum permeability $\mu_0$, and the volume-to-surface ratio of the magnetic field $H$. The corresponding $Q$ due to this surface resistance will be $Q_{\mathrm{int,H}} = G/R_\mathrm s$ \cite{creedon20163d}. In other words, the geometric factor is a measure of how much surface resistance is reduced due to the cavity's shape. Similarly to the oxide losses, geometric factor values can be calculated using the model as shown in Table 1. From the model, the geometric factors are in the range of several hundreds, which are typical values for three-dimensional cavities \cite{creedon20163d}. These are large values compared to two-dimensional resonators, where the geometrical values are typically below 1 \cite{zmuidzinas2012superconducting}.
\section{Effect of temperature on resonance shift}
\label{sec:resonance shift}
A shift in the fundamental frequency is observed as the cavity temperature goes from base temperature to around the superconducting transition temperature $T_\mathrm c$ as shown in Fig.~3a. This shift can be explained by the dependence of the kinetic inductance $L_\mathrm k$ on temperature through the material superconducting energy gap $\Delta$, which changes with temperature. For a resonator with an equivalent resistance $R$, capacitance $C$, and geometric inductance $L_0$, the resonance frequency is dependent on the kinetic inductance as  \cite{godse2020electronic}
\begin{equation}
f = \frac{1}{2\pi\sqrt{C(L_0+L_\mathrm k)}}.
\end{equation}
This can be approximated as
\begin{equation}
f = \frac{1}{2\pi\sqrt{CL_0}}(1-\frac{L_\mathrm k}{2L_0}),
\end{equation}
and the kinetic inductance is given by \cite{annunziata2010tunable}
\begin{equation}
L_\mathrm k = \frac{R\hbar}{\pi \Delta \tanh{\frac{\Delta}{2k_\mathrm B T}}}.
\end{equation}
In Bardeen–Cooper–Schrieffer (BCS) theory, the temperature dependence of $\Delta$ can be approximated as \cite{gross1986anomalous,poole2014superconductivity}
\begin{equation}
\Delta(T) = 1.76 k_\mathrm B T_\mathrm c \tanh{\frac{\pi}{1.76}\sqrt{\frac{T_\mathrm c}{T}-1}}.
\end{equation}
The fitting equation then becomes
\begin{equation}
f(T) = \delta + \frac{\epsilon}{E},
\end{equation}
where $E = \Delta \tanh{(\Delta/2k_\mathrm B T)}$, $\delta =  1/(2\pi \sqrt{CL_0})$, and $\epsilon = \hbar R/(4\pi^2 \sqrt{L_0^3 C}) $. Here, $\delta$ and $\epsilon$ are fitting parameters and $\hbar$ and $k_\mathrm B$ are Planck's and Boltzmann's contants, respectively. By setting $T_\mathrm c$ to 1.17\,K \cite{faber1955penetration}, the model produced a good fit as shown in Fig.~3a, which yields parameter values $\delta$ = 2.5123 GHz and $\epsilon$ = 2.28 $\times$ $10^{-19}$ Js\textsuperscript{-1}. Taking the ratio of these two fitting parameters gives $R/L_0$ = 5.4 $\times$ $10^{6}$ s\textsuperscript{-1}. More significantly, this behaviour of the resonance shift with respect to temperature very much resembles the form of the phenomenological expression $\Delta(T)$ in Eq.~6, which has almost no dependence when $T$ is close to zero and then makes a sharp and abrupt fall as $T$ approaches $T_\mathrm c$.
\section{Effect of temperature on quality factor}
\label{sec:qfactor shift}
Figure~3b shows how the quality factor of the fundamental mode changes as the temperature decreases from $T_\mathrm c$ to the base temperature. This behaviour of the quality factor can be seemingly described in two temperature ranges. One is from $T_\mathrm c$ to around 220\,mK and the other is from 220\,mK to the base temperature. In the first range, there is about a thousand fold change of the quality factor. On the other hand, the second range, while there is not much change in the value of $Q$, a small but noticeable peak occurs at about 130\,mK, where the quality factor reached about 27.6\,million. This two-range description seemingly hints to two temperature dependent loss mechanisms. As it turns out, there are two mechanisms that behave differently with temperature. These are the surface resistance predicted by the BCS theory \cite{turneaure1968microwave} and the losses due to two-level systems (TLS)\cite{phillips1987two}. These losses have different origins and therefore behave differently with temperature change. BCS losses come from the heating of unpaired electrons whose number increases with temperature. This resistance is dependent on temperature with the expression $\exp{(-\Delta/(k_\mathrm B T))}/T$ \cite{gurevich2017theory}. TLS losses, on the other hand, as the name suggests, stem from microscopic objects within the cavity wall that have two distinct energy levels, creating a quantum system where atoms can tunnel between these energy minima. These objects can be from any impurities, such as non-aluminium elements that are present in the alloy, or any other defects and dislocations in the lattice structure. TLS losses depend on temperature with the expression $\tanh{(\hbar \omega/(2 k_\mathrm B T))}$\cite{pappas2011two,romanenko2020three}. Using these two expression, we can create a model that can describe the behaviour of the quality factor over the aforementioned two temperature ranges as
\begin{eqnarray}
\frac{1}{Q_{\mathrm{total}}} & = \alpha \frac{1}{Q_{\mathrm{BCS}}} + \beta \frac{1}{Q_{\mathrm{TLS}}} + \gamma \frac{1}{Q_{\mathrm{residual}}} \nonumber \\
& = \alpha \frac{\exp{(\frac{-\Delta}{k_\mathrm B T})}}{T} + \beta \tanh{\frac{\hbar \omega}{2 k_\mathrm B T}} + \gamma,
\end{eqnarray}
where $\alpha$, $\beta$, $\gamma$ are the fitting factors for BCS losses, TLS losses, and any other residual losses that are not dependent on temperature, respectively. Using the same $T_\mathrm c$ = 1.17\,K for the resonance shift model, and $Q(0) = Q$ at base temperature, a reasonable fit was produced using this model shown in Fig.~3b, which yields parameter values $\alpha$ = 1.8795 $\times$ $10^{-6}$, $\beta$ = 3.1234 $\times$ $10^{-9} $, and $\gamma$ = 3.5211 $\times$ $10^{-8}$. To further understand the contributions from various loss mechanisms, we plot each weighted contribution separately as shown in Fig.~3b.

Around $T_\mathrm c$, it is clear that the dominant contribution comes from BCS losses, which are about 25 times larger than the residual losses. BCS losses steadily decrease as the cavity temperature approaches about 300\,mK and then sharply drop afterwards as the temperature approaches base temperature. TLS losses, on the other hand, increase by about 20 times gradually from $10^{-10}$ at $T_\mathrm c$ all the way down to base temperature. The aforementioned temperature range boundary at 220\,mK is the point, in fact, at which the crossing of the two loss contributions occurs. The combination of these two behaviours explains the peak observed at around 130\,mK. Interestingly, while BCS and TLS losses explain the behaviour of $Q$ with temperature, they are not the dominant contributor of losses in the experiment. Residual losses have the largest weight at about 92\,\% at base temperature. To understand the origin of this, a through measurement was done separately, replacing the cavity with a coaxial line to determine the coupling $Q$. From the through measurements, the coupling $Q$ is calculated to be 5 $\times$ $10^{8}$, leading to the unaccounted internal $Q$ to be 2.7532 $\times$ $10^{7}$. A possible source of this internal $Q$ is the loss from the presence of dielectric. However, as discussed earlier, even increasing either the oxide thickness from 3\,nm or introducing surface roughness to increase surface participation of the electrical field, this is not enough to account for this magnitude of loss. Another possible cause of loss could be the presence of the tuning rod holes. Firstly, energy from inside the cavity can leak out through these holes and, secondly, quasiparticles can be excited by external radiation entering the cavity through these holes. The former can be estimated by $\exp{(-4 \pi z/r)} $ for the fraction of power that is lost through a circular waveguide of radius $r$ and length $z$ when $d \ll \lambda$, where $\lambda$ is the radiation wavelength \cite{nikitin2008electromagnetic}. The estimated power reduction for the fundamental mode is about 5.2 $\times$ 10\textsuperscript{-39} for a $r$ = 3.183\,mm and $z$ = 22.5\,mm for the two holes, which is much lower than the measured residual losses. To verify whether quasiparticle excitation by the incoming radiation can explain the residual losses, we repeated the experiment with both holes patched by an adhesive copper foil with a conductive glue. The fact that the $Q$ factor did not change rules out this effect as a possible cause. Hence, the residual losses are most likely caused by the presence of impurities in the cavity material. According to the alloy specifications \cite{kaufman2000introduction}, the material is composed of 95.85\,\% aluminium and a few other elements, the dominant impurities being magnesium (1.0\,\%) and silicon (0.6\,\%). Understanding of exact loss mechanisms caused by different impurities can be a topic of further studies.
\section{Conclusion}
In summary, we have characterised an aluminium cylindrical cavity at its superconducting temperature range. Using finite element method simulations, we were able to understand how the geometry of three-dimensional cavities results in very low surface participation of both dielectric losses and surface resistance, leading to theoretically extremely high $Q$ values. In practice, understanding the effect of temperature on $Q$ starts with the seeing the effect of the temperature on the superconducting energy gap. This effect is clearly seen in the sharp resonance frequency shift with temperature near $T_\mathrm c$, where the change in the superconducting energy gap affects the kinetic inductance of the cavity. Finally, the temperature dependence of $Q$ can be described as a combination of BCS and TLS losses, where BCS contributions dominate near $T_\mathrm c$ and TLS losses become more prominent at temperatures near zero. This mixture of behaviour results in a small peak in $Q$. Remarkably, through this experiment we learned that we can achieve a high internal $Q$ even with a cavity material with a lot of impurities. This means that by using three dimensional superconducting cavities with smooth wall transitions such as this cavity, it allows you to have the best ceiling of $Q$ values while also having the best baseline of low losses due to surface imperfections as compared to cavities of other geometries. In the context of light dark matter searches, type I superconducting cavities such at the one characterised in this study can be deployed in the search for paraphotons where an external magnetic field is not required, and the signal power directly scales with the quality factor of the cavity. For axion searches, such a cavity may still be suitable if higher order modes of the cavity are exploited and a heterodyne approach is used \cite{berlin2021heterodyne} such that the magnetic field is provided by the resonant modes of the cavity itself. In earlier studies of axion haloscopes \cite{sikivie1985detection} it was concluded that increasing the quality factor of the cavity much about $10^6$ would be of limited advantage due to the effective quality factor of the axion field itself originating in its velocity spread. More recent studies \cite{kim2020revisiting}, however, show that there is still an advantage in improving the cavity quality factor to the levels demonstrated in this paper. Overall, this analysis and understanding of these loss mechanisms will be important to any area this technology is applied.

\ack
We thank the ADMX group members who designed the cavity and Eroda Ltd \cite{Eroda}. who manufactured it. We also thank Sam Furness for his contributions to the measurements and data analysis. The Lancaster and Sheffield groups acknowledge the support of the Science and Technology Facilities Council (STFC) with the Quantum Sensors for the Hidden Sector (QSHS) project (ST/T006102/1), ParaPara project (ST/W006502/1), and the European Research Council (grant 824109). The LLNL group is supported by the U.S. Department of Energy under Contract DE-AC52-07NA27344. LLNL-JRNL-XXXXXX.

\section*{References}
\providecommand{\newblock}{}

\newpage

\begin{figure}[p]
    \centering
    \includegraphics[width=1\columnwidth]{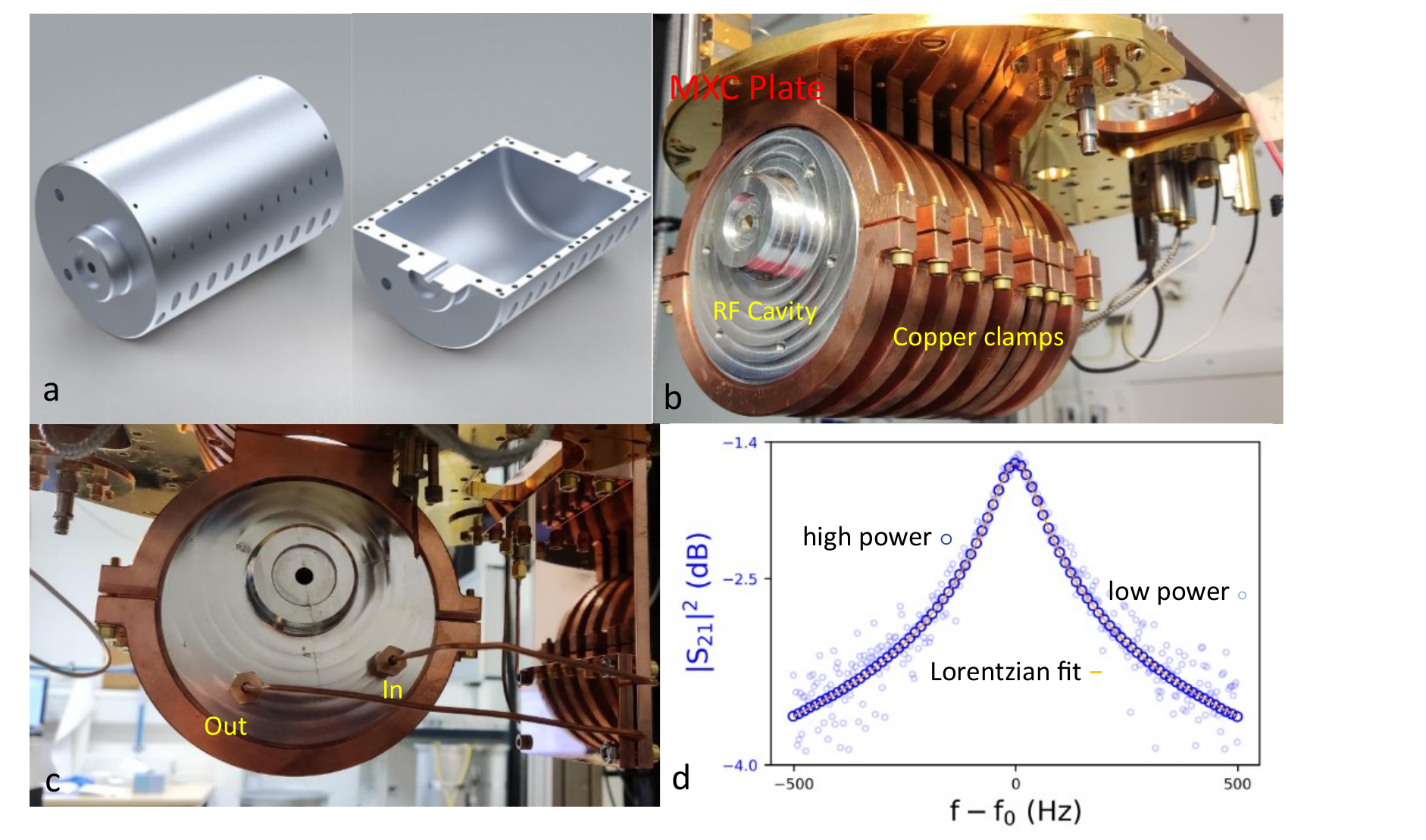}
    \caption{(a) Renders of the measured cavity. (b) Aluminium cylindrical cavity clamped to the mixing chamber plate using copper brackets tightened with brass screws. (c) Copper feedthroughs are secured to the cavity using collet screws. (d) Transmission of the TM\textsubscript{010} mode of the cavity with a high and low power input and corresponding Lorentzian fit.}
    \label{fig:Aluminium Cylindrical Cavity}
\end{figure}

\newpage

\begin{figure}[p]
    \centering
    \includegraphics[width=1\columnwidth]{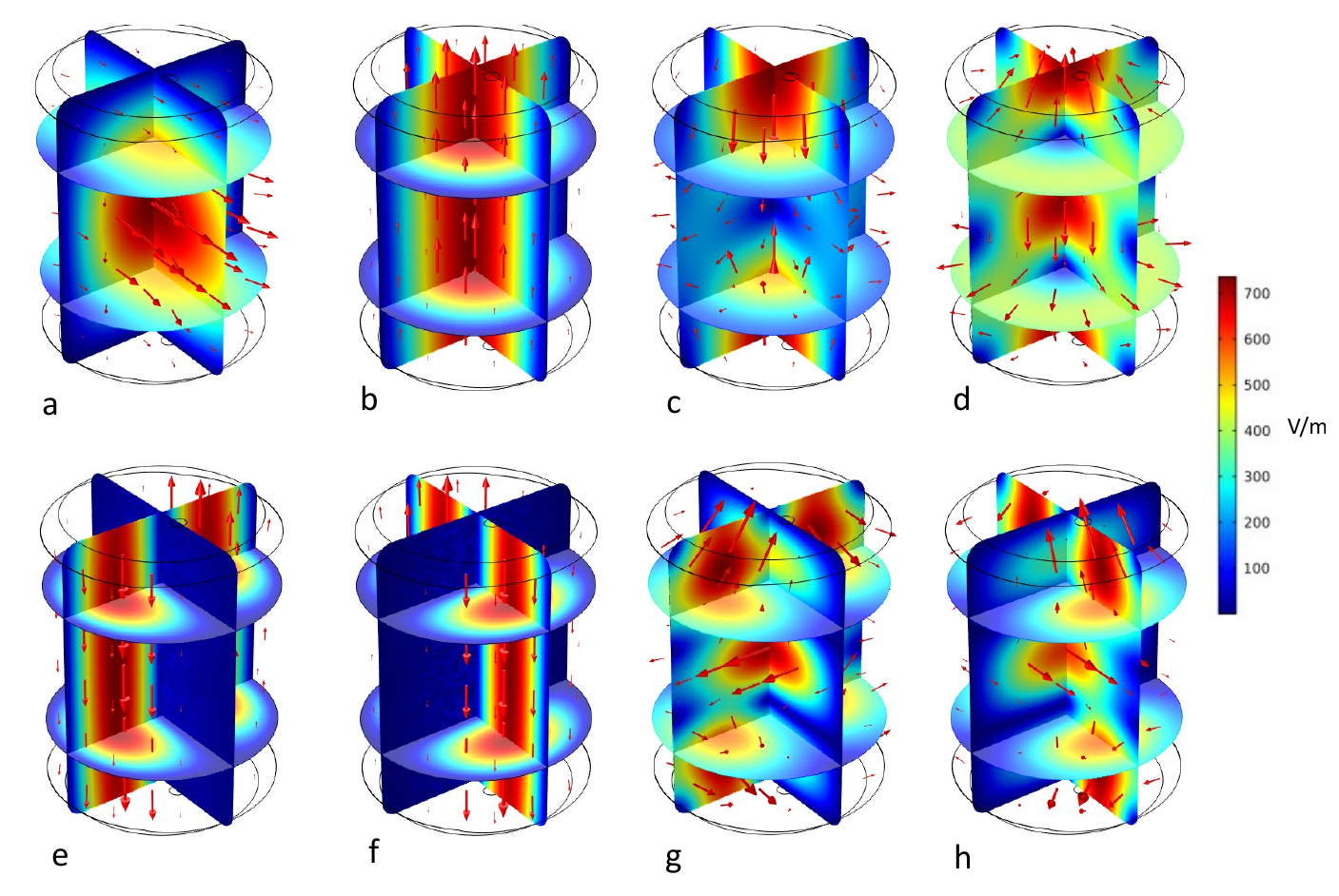}
    \caption{Visualisation of the different modes from the model, where the slices show the electric field magnitude heat map and the arrows indicate the electric field direction.}
    \label{fig:COMSOL models}
\end{figure}

\newpage

\begin{figure}[p]
    \centering
    \includegraphics[width=1.0\columnwidth]{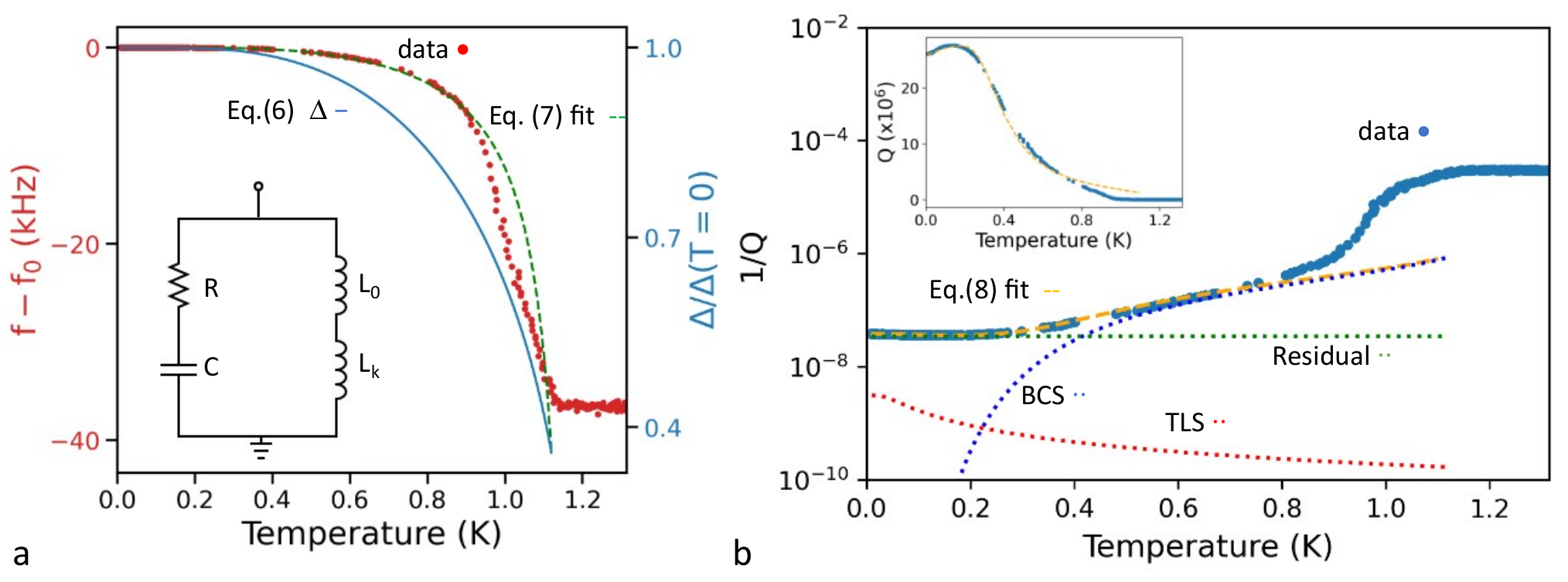}
    \caption{(a) Resonance shift of the TM\textsubscript{010} mode as a function of cavity temperature and equivalent circuit (inset). (b) Temperature dependence of the quality factor of the TM\textsubscript{010} mode (inset) and the weighted contributions of various loss mechanisms to the total loss.}
    \label{fig:Resonance Shift}
\end{figure}

\newpage

\begin{table*}[p]
\begin{centering}
\begin{tabular}{|c|c|c|c|c|c|}
\hline 
Mode & Measured & Measured & Simulated & Dielectric$^\dagger$&Geometric\\
 & Freq. (GHz) &  $Q$ Factor ($\times$ 10$^6$) & Freq. (GHz) & Participation $P_{\mathrm{diel}}$ &Factor $G$ ($\Omega$)\\
\hline
\hline 
TE\textsubscript{101} & N/A & N/A & 2.2866 & 5.48e-7 & 309\tabularnewline
\hline 
TM\textsubscript{010} & 2.5123 & 26.095 & 2.5118 & 4.76e-7 & 342\tabularnewline
\hline 
TM\textsubscript{011} & 2.8156 & 22.086 & 2.8155 & 1.03e-6 & 306\tabularnewline
\hline 
TM\textsubscript{012} & 3.5571 & 23.599 & 3.5548 & 1.14e-6 & 384\tabularnewline
\hline 
TM\textsubscript{110a} & 4.0007 & 24.730 & 4.0012 & 4.44e-7 & 551\tabularnewline
\hline 
TM\textsubscript{110b} & 4.0025 & 0.066 & 4.0012 & 4.47e-7 & 550\tabularnewline
\hline 
TM\textsubscript{111a} & 4.2245 & 25.438 & 4.2232 & 6.74e-7 & 582\tabularnewline
\hline 
TM\textsubscript{111b} & 4.2248 & 8.892 & 4.2232 & 7.58e-7 & 546\tabularnewline
\hline 

\end{tabular}
\par\end{centering}
\caption{Extracted cavity values for the different modes from the experiment and simulations
\newline Notes: $\dagger$ Estimate for an oxide layer of thickness $t$ = 3\,nm}
\end{table*}

\end{document}